\pdfoutput=1

\documentclass[preprint2]{aastex}

\shorttitle{Parallax and orbital effects in astrometric microlensing with binary sources}
\shortauthors{Nucita et al.}

\usepackage[utf8]{inputenc}
\usepackage{graphicx,epsfig}  
\usepackage{amsmath,amssymb}
\usepackage{subfigure}
\usepackage{longtable}
\usepackage{lscape}
\usepackage{times}
\usepackage{textcomp}
\usepackage{natbib}
\usepackage{multirow}

\def\ut#1{\mathop{\vtop{\ialign{##\crcr
     $\hfil\displaystyle{#1}\hfil$\crcr\noalign
     {\kern1pt\nointerlineskip}\hbox{$\hfil\sim\hfil$}\crcr
     \noalign{\kern1pt}}}}}

\def\undersymbol#1#2{\mathop{\vtop{\ialign{##\crcr
     $\hfil\displaystyle{#2}\hfil$\crcr\noalign
     {\kern1pt\nointerlineskip}\hbox{$\hfil#1\hfil$}\crcr
     \noalign{\kern1pt}}}}}

\def\degr{^\circ}

\begin{document}

\title{Parallax and orbital effects in astrometric microlensing with binary sources}
\author{A. A. Nucita\altaffilmark{1,2}, F. De Paolis\altaffilmark{1,2}, G. Ingrosso\altaffilmark{1,2}, M. Giordano\altaffilmark{1,2}, and L. Manni\altaffilmark{1,2}}
\affil{Department of Mathematics and Physics {\it `E. De Giorgi'}, University of Salento, Via per Arnesano, CP 193, I-73100,
Lecce, Italy}
\affil{INFN, Sez. di Lecce, Via per Arnesano, CP 193, I-73100, Lecce, Italy}
\email{nucita@le.infn.it}

\label{firstpage}
\begin{abstract}
{  In gravitational microlensing, binary systems may act as lenses or sources. Identifying lens binarity is generally easy especially in events 
characterized by caustic crossing since the resulting light curve exhibits strong deviations from smooth single-lensing light curve. 
On the contrary, light curves with minor deviations from a Paczy\'nski behaviour do not allow one to identify the source binarity. A 
consequence of the gravitational microlensing is the shift of the position of the multiple image centroid with respect to the source star location - the so called astrometric microlensing signal. 
When the astrometric signal is considered, the presence of a binary source manifests with a path that largely differs from that expected 
for single-source events. Here, we investigate the astrometric signatures of binary sources taking into account their orbital 
motion and the parallax effect due to the Earth motion, which turn out not to be negligible in most cases. 
We also show that considering the above-mentioned effects is important in the analysis of
astrometric data in order to correctly estimate the lens-event parameters.}
\end{abstract}


\section{Introduction}

Gravitational microlensing is {  a} mature
technique for detecting compact objects in the disk and in the halo of our Galaxy
via the observation of the light magnification of source stars due to the intervening lenses. 
Indeed, the technological instrument advances allowed
gravitational microlensing {  to detect and characterized low-mass objects
(see e.g.  \citealt{park2015}) as well as
binary lens systems (see e.g. \citealt{udalski2015}) including
planetary systems with planets masses down to 
Earth mass with host-planet separations of about a few AU.}

{  In addition to the magnification of the source brightness, another}
phenomenon related to microlensing is the shift {  of the light 
centroid of the source images.} This subject was studied by many authors 
(see e.g. \citealt{walker1995}, \citealt{miyamoto1995}, \citealt{hog1995},
\citealt{jeong1995}, \citealt{paczinsky1996}, \citealt{paczinsky1998}, \citealt{dominik2000}, 
\citealt{takahashi}, \citealt{lee2010}). 
In the simplest case of a {  point lens, lensing causes the source image to split into two and} the position of 
the light centroid with respect to the {  unlensed} source star position traces out an ellipse 
with semi-axes {  depending}, in general, on the lens impact parameter $u_0$ (the minimum projected distance of the lens to the source star) {  and the shape of the astrometric trajectory does 
not depend on the Einstein time $t_E$.} 

When the lens is a binary system (see e.g. \citealt{han1999}, \citealt{griest1998}, \citealt{han2001},
\citealt{bozza2001}, \citealt{hideki}, \citealt{sajadian2015a}), the number and the position
of the images {  differ from those of the single lens case and the astrometric signal 
trajectory and the deviation varies depending on} on the binary system parameters (i.e., the mass ratio 
and the component separation).
 
It is evident that in both cases astrometry gives more information than that derived {  from} 
the analysis of {  light curves} (photometry), allowing one to {  better constrain the lens system}.
\footnote{  We mention that other methods to face the parameter 
degeneracy problem rely on the measurement of the 
lens proper motion (see e.g. \citealt{propermotion}) or on polarization observations 
\citep{ingrossoa,ingrossob} in ongoing microlensing events.}

A further advantage of the astrometric microlensing is that an event is potentially observable 
for a much longer time with respect to the typical photometric event {  because astrometric signals persit to much longer lens-source separations than photometric signals} (see next sections). 
In addition, interesting events can be 
predicted in advance (\citealt{paczinsky1995}) and, indeed, by studying in detail the characteristics of stars with large proper 
motions, \citet{proft2011} identified {  dozens} of candidates {  for astrometric 
microlensing observations using the Gaia satellite}, an European Space Agency (ESA) mission, 
that is performing photometry, spectroscopy and high precision astrometry (\citealt{eyer}). 

Binary star systems {  can} act as sources of microlensing events. In this {  regard}, each component of
the binary system acts as {  an independent source} (with given impact parameter) for the intervening lens 
and the {  resultinng light curve corresponds to a a superposition of the single-lensing 
light curves associated with the individual source stars}. However, although \citet{griest1992} predicted that about $10\%$ of the observable 
events should involve features of a binary source, {  few} clear detection of such systems was claimed 
up today\footnote{\citet{jaroszynski2004}, analyzing the OGLE-III Early Warning System database for 
seasons 2003-2004, reported 15 events possibly interpreted as binary sources lensed by single objects (see also \citealt{hwang2013}).}.
As argued by \citet{dominik1998b}, the lack of binary {  source events}  
may be explained by the fact that most of the light curves for events involving a {  binary source can be explained by
single lens model with a blended source}. So, binary sources are hidden in photometric observations.
This  is certainly  not the case for astrometric microlensing observations, 
for which, as first {  pointed out} by \citet{hankim1999} 
and \citet{dalalgriest2001}, the binarity of the source strongly modifies {  
astrometric signals}.  However, these authors accounted for the binary source effect 
by considering the centroid shift as due mainly to the primary object while treating its companion 
as a simple blending source. This simplifying assumption is overcome in the present paper where 
both components of the binary source and their relative motion 
are considered in calculating the resulting astrometric path. 

Several theoretical studies (see e.g. \citealt{dominik1997, dominik1998a}, 
\citealt{penny2011a,penny2011b}, \citealt{nucita2014}, \citealt{giordano2015}) already pointed out the 
importance of considering the orbital motion of a binary lens system in microlensing {  light curves} and 
observation of peculiar microlensing events (see e.g. \citealt{park2015}, \citealt{Skowron2015}, and \citealt{udalski2015}){ , 
demonstrating the necessity to account for such effect. Here, we investigate the effects on the astrometric signals of the binary source orbital motion 
taking also into account the Earth parallax effect. We show that both effects are not negligible 
in most astrometric microlensing observation.}

The paper is structured as follows: in Section 2, we briefly review the basics of astrometric 
microlensing for a single lens and source. In Section 3 we discuss the expected astrometric signal for binary source events (static or not) lensed  
by single or binary objects and show that the centroid shift trajectories strongly deviate from the pure elliptical shape. In Section 4, we consider the Earth motion 
and study the deviation in astrometric curves induced by the parallax effect. We address our conclusion in Section 5.

\section{Basics of astrometric microlensing}
For a source at angular distance $\theta_S$
from a point-like gravitational lens, the {  positions} $\theta$ of the images with respect
to the lens {  are obtained} by solving the lens equation \citep{falco1992}
\begin{equation}
\theta^2-\theta_S\theta -\theta_E^2=0,
\end{equation}
where $\theta_E$ is the Einstein angle 
\begin{equation}
\theta_E=\left(\frac{4GM}{c^2}\frac{D_{S}-D_L}{D_L D_S}\right)^{\frac{1}{2}},
\end{equation}
being $M$ the lens mass, $D_{S}$ and $D_L$ the distances from the observer to the source and lens, respectively.
{  When the Einstein radius is expressed as a linear scale $R_E=D_L\theta_E$ the lens equation becomes}
\begin{equation}
d^2-d_S d -R_E^2=0,
\end{equation}
where $d_S$ and $d$ are the linear distances (in the lens plane) of the source and images from the gravitational lens, respectively.
Using the dimensionless quantities
\begin{equation}
u=\frac{\theta_s}{\theta_E},~~~ \tilde{u}=\frac{\theta}{\theta_E},
\end{equation}
the lens equation can be further simplified as
\begin{equation}
\tilde{u}^2-u\tilde{u} -1=0.
\end{equation}
The solutions of this equation
\begin{equation}
\tilde{u}_{+,-}=\frac{1}{2}\left[u\pm\sqrt{4+u^2}\right],
\end{equation}
give the locations of the positive and negative parity images ($+$ and $-$, respectively) with respect to 
the lens position. The two images have magnifications
\begin{equation}
\mu_{+,-}=\frac{1}{2}\left[1\pm\frac{2+u^2}{u\sqrt{4+u^2}}\right],
\end{equation}
{  so that the} total magnification is (\citealt{paczinsky1986}),
\begin{equation}
\mu=|\mu_+|+|\mu_-|=\frac{2+u^2}{u\sqrt{4+u^2}}.
\label{amplification}
\end{equation}
Note that, in the lens plane, the $+$ image resides always outside a circular ring centered on the lens position with radius equal to the Einstein angle,
while the $-$ image is always within the ring. As the source-lens distance increases, the $+$ image approaches the source position while the $-$ one (becoming fainter) moves towards the lens location. For $u \ll 1$,
the magnification can be approximated\footnote{Considering 
the next order approximation, one gets
\begin{equation}
\mu\simeq \frac{1}{u}+\frac{3u}{8},~~~{\rm for}~u~\ll~1,
\end{equation}
and
\begin{equation}
\mu\simeq 1+\frac{2}{u^4}-\frac{8}{u^6},~~~{\rm for}~u~\gg~1.
\end{equation}
} as (see e.g. \citealt{dominik2000})
\begin{equation}
\mu\simeq \frac{1}{u},
\end{equation}
while for $u \gg 1$, one has
\begin{equation}
\mu\simeq 1+\frac{2}{u^4},
\label{magfalls}
\end{equation}
so that for large angular separations, the lensing effect produces a source magnitude shift of
\begin{equation}
\Delta m\simeq -\frac{5}{\ln 10 ~u^4}.
\end{equation}

Let us consider now a source moving in the lens plane with transverse velocity $v_{\perp}$ and let be $\xi L \eta$ a frame of reference centered on the lens, with the $\xi$ axis oriented along the 
velocity vector and $\eta$ axis perpendicular to it. {  Then}, the projected coordinates of the source 
(in units of the Einstein radius) result to be
\begin{equation}
\xi(t)=\displaystyle{\frac{t-t_0}{t_E}},~~~\eta(t)=\displaystyle{u_0}, 
\label{eqcoordinatescenter}
\end{equation}
where $t_E=R_E/v_{\perp}$ is {  the Einstein time scale} of the event and 
$u_0$ is the distance of closest approach  or impact parameter (in this case lying on the $\eta$ axis) occurring at time $t_0$. 
Thus, since $u^2=\xi^2+\eta^2$, the two images move in the lens plane during the gravitational lensing event.
The centroid of the image pair can be defined as the average position of the $+$ and $-$ images weighted {  by the associated magnifications} (\citealt{walker1995})
\begin{equation}
\bar{u}\equiv\frac{\tilde{u}_+\mu_+ +\tilde{u}_-\mu_-}{\mu_+ +\mu_-}=\frac{u(u^2+3)}{u^2+2},
\label{centroidpair}
\end{equation}
so that, by symmetry, the image centroid is always at the same azimuth as the source. The measurable quantity is the displacement of the centroid of the image pair relative to the source, i.e.
\begin{equation}
\Delta\equiv\bar{u}-u=\frac{u}{2+u^2},
\label{shiftmodulus}
\end{equation}
{  which is a function of the time} since $u$ is time dependent. One can easily realize that $ {\Delta}$ may be viewed 
as a vector 
\begin{equation}
{  \Delta}=\frac{{  u}}{2+u^2}
\label{shiftarray}
\end{equation}
with components along the axes
\begin{equation}
\Delta_{\xi}(t)=\displaystyle{\frac{\xi(t)}{2+u(t)^2}},~~~ \Delta_{\eta}(t)=\displaystyle{\frac{\eta(t)}{2+u(t)^2}}.
\label{shiftcomponents}
\end{equation}
Here, we remark that all the angular quantities 
are given in units of the Einstein angle $\theta_E$ {  which is related to the physical lens parameters as}
\begin{equation}
\theta_E\simeq2\left(\frac{M}{0.5 {\rm ~M_{\odot}} }\right)^{1/2}\left(\frac{D_L}{{\rm kpc} }\right)^{-1/2}~{\rm mas}.
\end{equation}
Note that while the $\Delta_{\eta}$ component is symmetric with respect to $t_0$ and always positive,
the $\Delta_{\xi}$ component is an anti-symmetric function with minimum and maximum values occurring 
at $t_0 \pm t_E \sqrt{u_0^2+2}$, respectively.

One can also verify that, in contrast to the magnification $\mu$ 
(which diverges for $u_0 \rightarrow 0$), {  the maximum centroid shift equals to $\sqrt{2}/4$ for $u_0=\sqrt{2}$.}
In particular, due to the anti-symmetry of the $\xi$ component, for $u_0<\sqrt{2}$ the shift goes 
through a minimum at $t=t_0$ and has two maxima at $t_0\pm t_E\sqrt{2-u_0^2}$. Conversely, 
for $u_0 \ge \sqrt{2}$, $\Delta$ assumes the single maximum 
value equal to $u_0/(u_0^2+2)$ at $t=t_0$.
As first noted by \citet{dominik2000}, for $u\ll \sqrt{2}$ the centroid shift tends 
linearly to zero (hence, $\Delta \simeq  u/2$) while the photometric magnification increases 
towards small lens-star separation. In addition, for $u\gg \sqrt{2}$ one has 
$\Delta \simeq 1/ u$, so that the centroid shift
falls more slowly than the magnification -- see eq. (\ref{magfalls}) --
thus implying that the centroid shift could be a {  promising} observable also 
for large source-lens distances, i.e.
far from the light curve peak. In fact, 
in astrometric microlensing the threshold impact parameter $u_{\rm th}$ (i.e. the value of 
the impact parameter that gives an astrometric centroid signal larger than a certain 
quantity $\delta_{\rm th}$) 
is $u_{\rm th}=\sqrt{T_{\rm obs} v_{\perp}/(\delta_{\rm th} D_L)}$ where $T_{\rm obs}$ 
is the observing 
time and $v_{\perp}$ the relative velocity of the source 
with respect to the lens. {  For example,} the
Gaia satellite would reach an astrometric precision $\sigma_G\simeq 300$ $\mu$as 
(for objects with visual magnitude $\simeq 20$) in $5$ years of observation 
(\citealt{eyer}). Then, assuming a threshold centroid shift $\delta_{\rm th}\simeq \sigma_G$,
one has $u_{\rm th}\simeq 60$ for $D_L=0.1$ 
kpc and $v_{\perp} \simeq 100$ km s$^{-1}$. 
For comparison, the threshold impact parameter for a ground-based photometric observation 
is $\simeq 1$. Thus, the cross section for astrometric microlensing, and consequently the event rate, 
is much larger than {  that of} the photometric observation since it scales as $u_{\rm th}^2$.

It is straightforward to show (\citealt{walker1995}) that during a microlensing event
the centroid shift $\Delta$ traces (in the 
$\Delta_{\xi}, \Delta_{\eta}$ plane) an ellipse centered in the point 
$(0,b)$. The ellipse semi-major axis $a$ (along $\Delta_{\eta}$) and 
semi-minor axis $b$ (along $\Delta_{\xi}$) are
\begin{equation}
a=\frac{1}{2}\frac{1}{\sqrt{u_0^2+2}},~~~b=\frac{1}{2}\frac{u_0}{u_0^2+2},
\label{axes}
\end{equation}
being evident that for $u_0 \rightarrow \infty$ the ellipse becomes a circle with radius $1/(2u_0)$ 
and it {  becomes} a straight line of length $1/\sqrt{2}$, for $u_0$ approaching zero. 
Note also that from eq. (\ref{axes}) one finds that
\begin{equation}
u_0^2=2 (b/a)^2 \left[1-(b/a)^2\right]^{-1}.
\label{u0fromaxes}
\end{equation}
Hence (in the absence of finite-source and blending effects)
by measuring $a$ and $b$, one can directly estimate
the impact parameter $u_0$. 
In addition, in the case of small impact parameters ($u_0 < \sqrt{2}$) the Einstein time $t_E$ can 
be readily derived by measuring the time lag between the peak features (see e.g. Figure 1 in \citealt{dominik2000}).

\section{Astrometric microlensing for a binary source}

Here, we study the astrometric path for a rotating binary source lensed by a single lens or by a binary system. 
As {  pointed out} by \citet{dominik1998b}, in the case of a binary 
source with a single intervening lens, the resulting light curve is 
the superposition of the Paczy\'{n}sky amplifications associated {  to the individual} binary components. Since, typically, only one source is highly magnified, 
the convolved light curve {  can be} well fitted by a single lens model with a blended source so that the binary source event is missed completely. 
However, as noted by \citet{hankim1999} (but see also \citealt{dalalgriest2001}), for binary source events the astrometric signal strongly deviates from that 
expected in the single source case. In particular, \citet{han2001astrometry} showed that the centroid shift at time $t$ can be obtained via a weighted 
average {  of} the individual source component amplifications 
{  with respect to the reference position} the centre of light between the unlensed source components, i.e. 
\begin{equation}
{  \Delta _{bs}} =\frac{\mu_1F_1({  u_1}+{  \Delta_1})+\mu_2F_2({  u_2}+{  \Delta_2})}{\mu_1F_1+\mu_2F_2}
-\frac{F_1{  u_1}+F_2{  u_2}}{F_1+F_2},
\label{deltatotalbs}
\end{equation}
where ${  u_i}$ are the distances between the lens and the individual binary source components, $\mu_i$ and ${  \Delta_i}$ the magnification factors and the centroid shifts of the 
two single sources (as given by eqs. \ref{amplification} and \ref{shiftarray}) having luminosity $F_i$ with subscripts $i=1$ and $i=2$ for the primary object and its companion, respectively.

Several studies (\citealt{dominik1997}, \citealt{penny2011a,penny2011b}, \citealt{nucita2014}, \citealt{giordano2015} and \citealt{penny2015}) and 
microlensing observations (\citealt{park2015}, \citealt{Skowron2015}, and \citealt{udalski2015}) pointed out the necessity to consider the orbital motion of the lens components in photometric studies. 

In astrometric observation of microlensing events, the lens orbital motion gives rise to single or multiple twists in the astrometric path of $ {\Delta}$ showing the importance of considering this effect in any 
fit procedure. The same is also true if one considers the astrometric signal due to binary sources. Let us {  define} by $m_1$ and $m_2$ {  as} the masses of the two source components 
(with $m_2<m_1$ so that $q=m_2/m_1<1$), and total mass normalized to unity, i.e. $m_1+m_2=1$. In this case, the {  separations of the individual source components from the 
center of mass are, respectively}
\begin{equation}
r_1=-\frac{\mu_r}{m_1}b,~~~~r_2=\frac{\mu_r}{m_2}b,
\end{equation}
where the reduced mass is $\mu_r = q/(1+q)^{2}$ and $b$ represents the binary semi-major axis in units of the $R_E$. Hence, 
the components of the position vectors of the binary source objects in the lens plane with respect to the lens (at the origin of the adopted 
reference frame) are 
\begin{equation}
\begin{split}
\xi_1(t)=\xi_{cm}(t)+r_1\cos \theta (t),~~\eta_1(t)=\eta_{cm}(t)+r_1\sin \theta (t), \\
\xi_2(t)=\xi_{cm}(t)+r_2\cos \theta (t),~~\eta_2(t)=\eta_{cm}(t)+r_2\sin \theta (t),
\end{split}
\end{equation}
where $\xi_{cm}(t)$ and $\eta_{cm}(t)$ are the coordinates at time $t$ of the center of mass, given in eq. (\ref{eqcoordinatescenter}) and the polar angle depends on the Keplerian orbital period $P$ as 
$\theta=2\pi(t-t_0)/P$. Note that in this toy-model we are assuming binary sources moving on circular orbits: the most general case 
of elliptic orbits (with $r_1$ and $r_2$ depending also on time $t$) can be easily accounted for by solving the associated Kepler problem 
(see e.g. \citealt{nucita2014} and references therein). 

\begin{figure*}
\centering
        \subfigure{%
            \includegraphics[width=0.9\columnwidth]{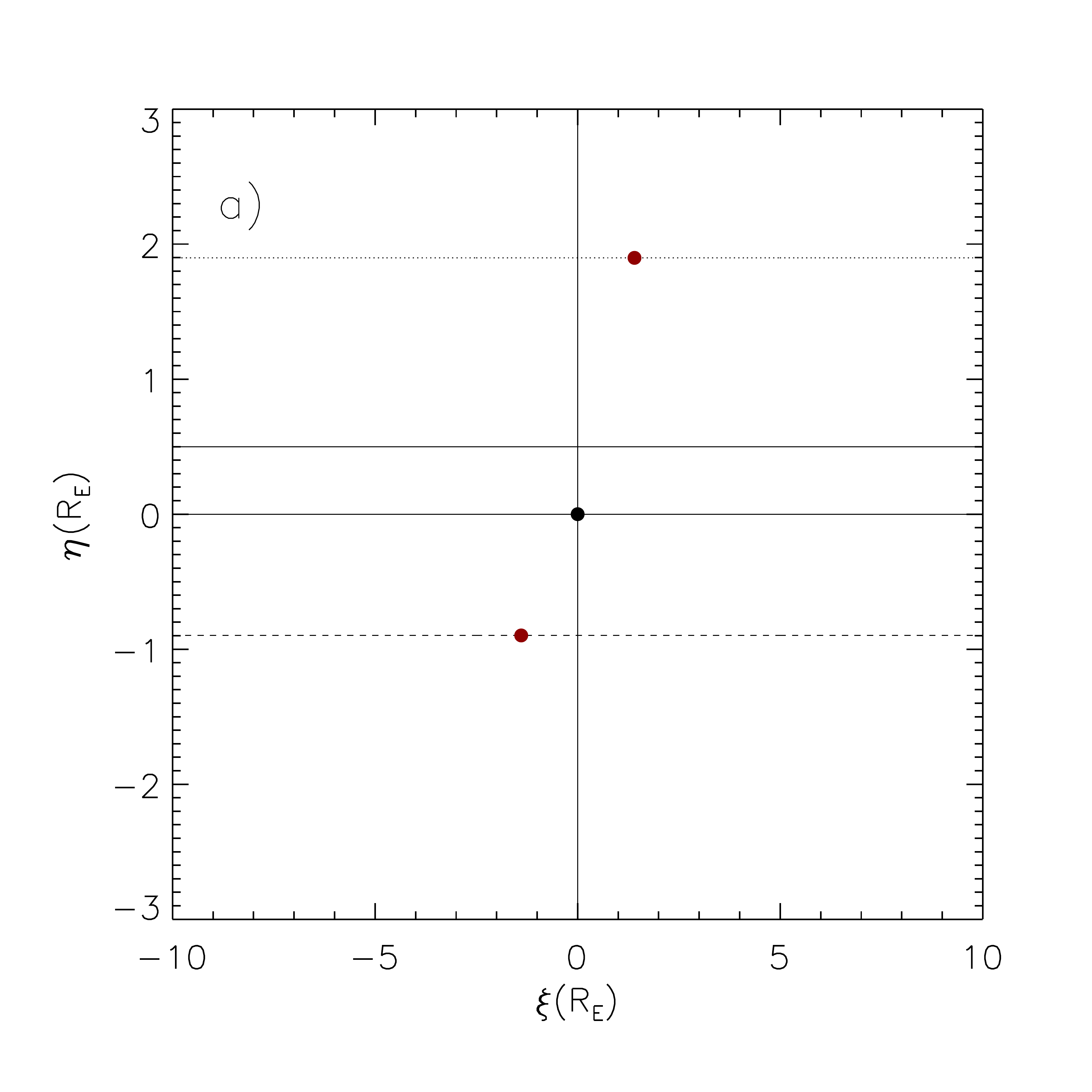}
        }
        \subfigure{%
           \includegraphics[width=0.9\columnwidth]{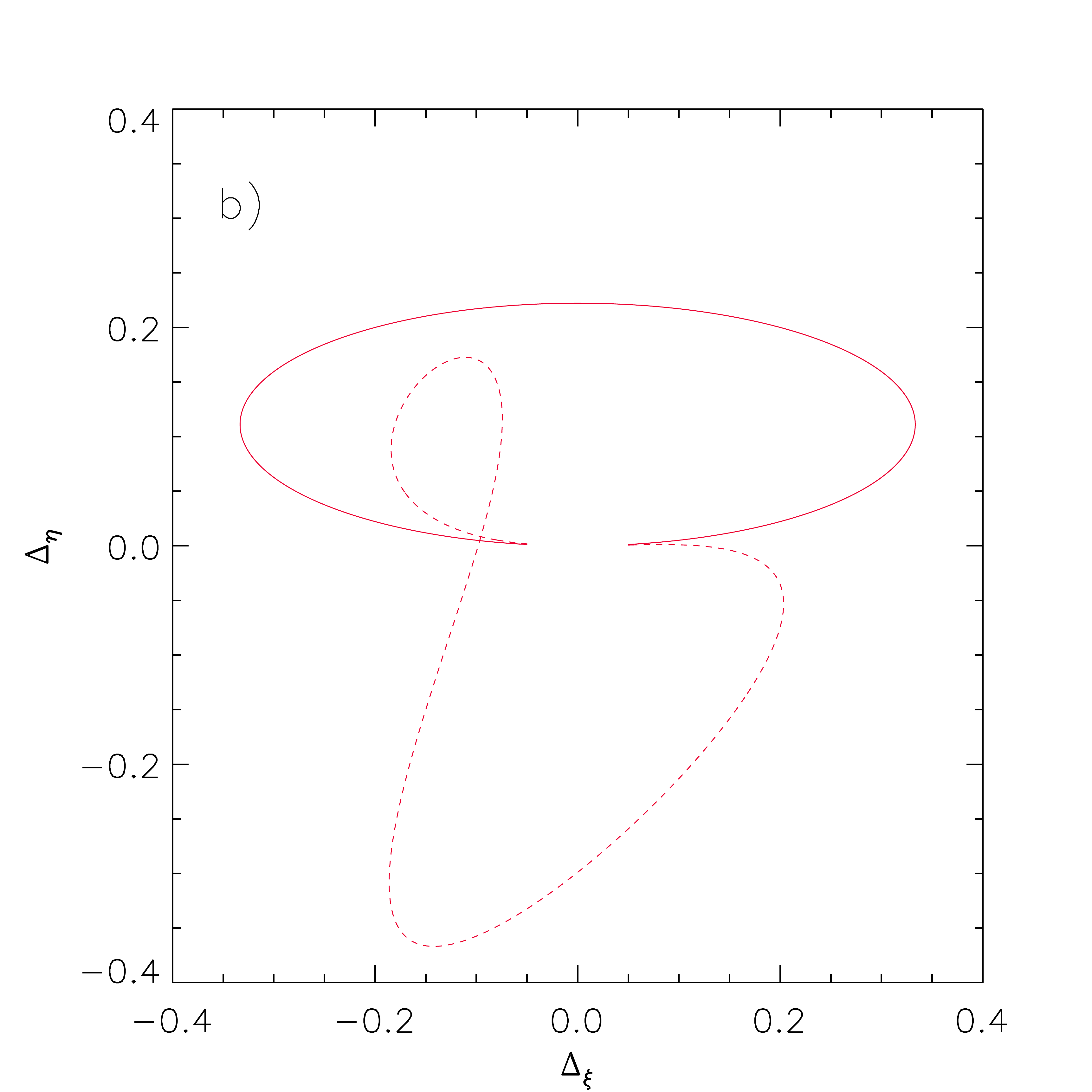}
        }\\
        \subfigure{%
            \includegraphics[width=0.9\columnwidth]{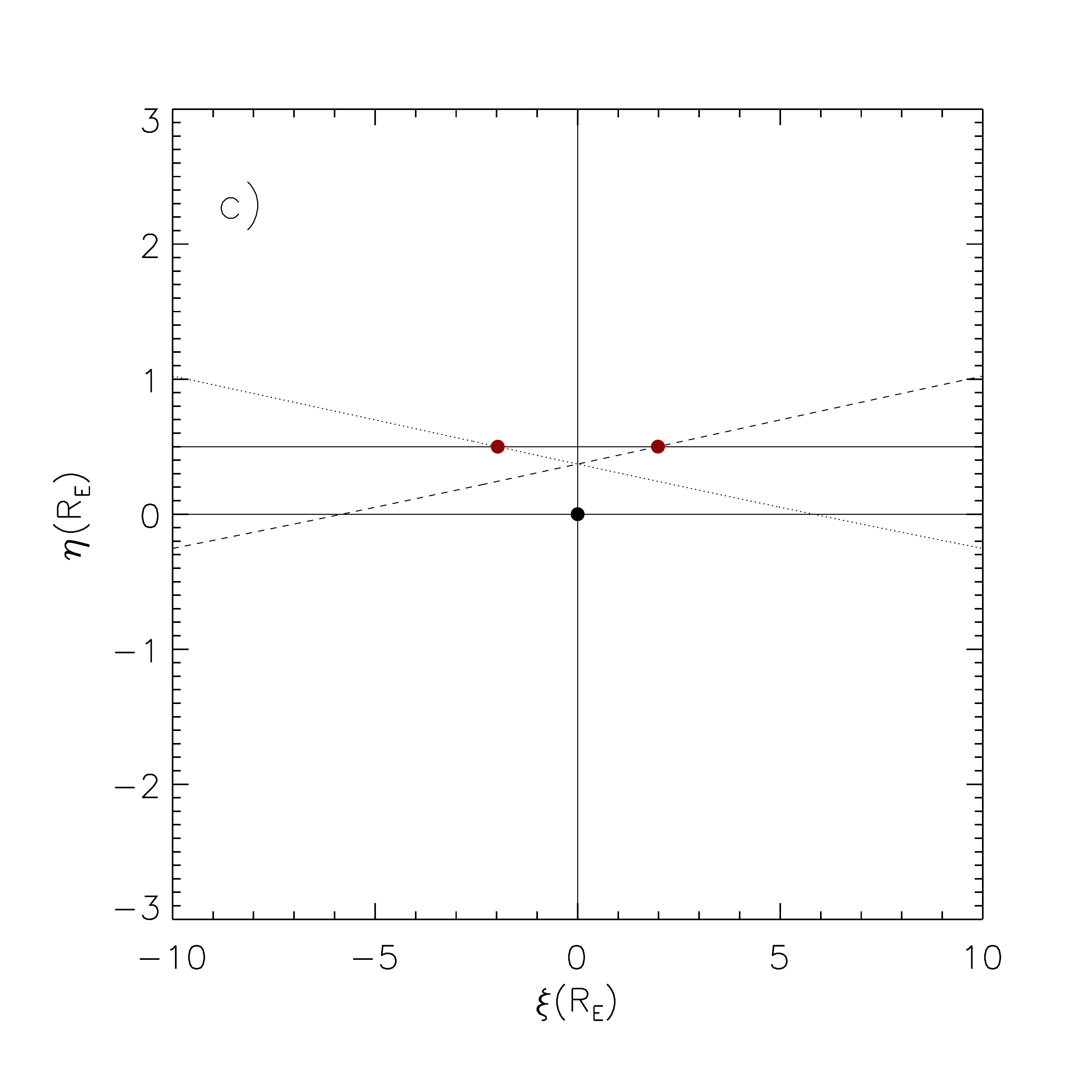}
        }%
        \subfigure{%
            \includegraphics[width=0.9\columnwidth]{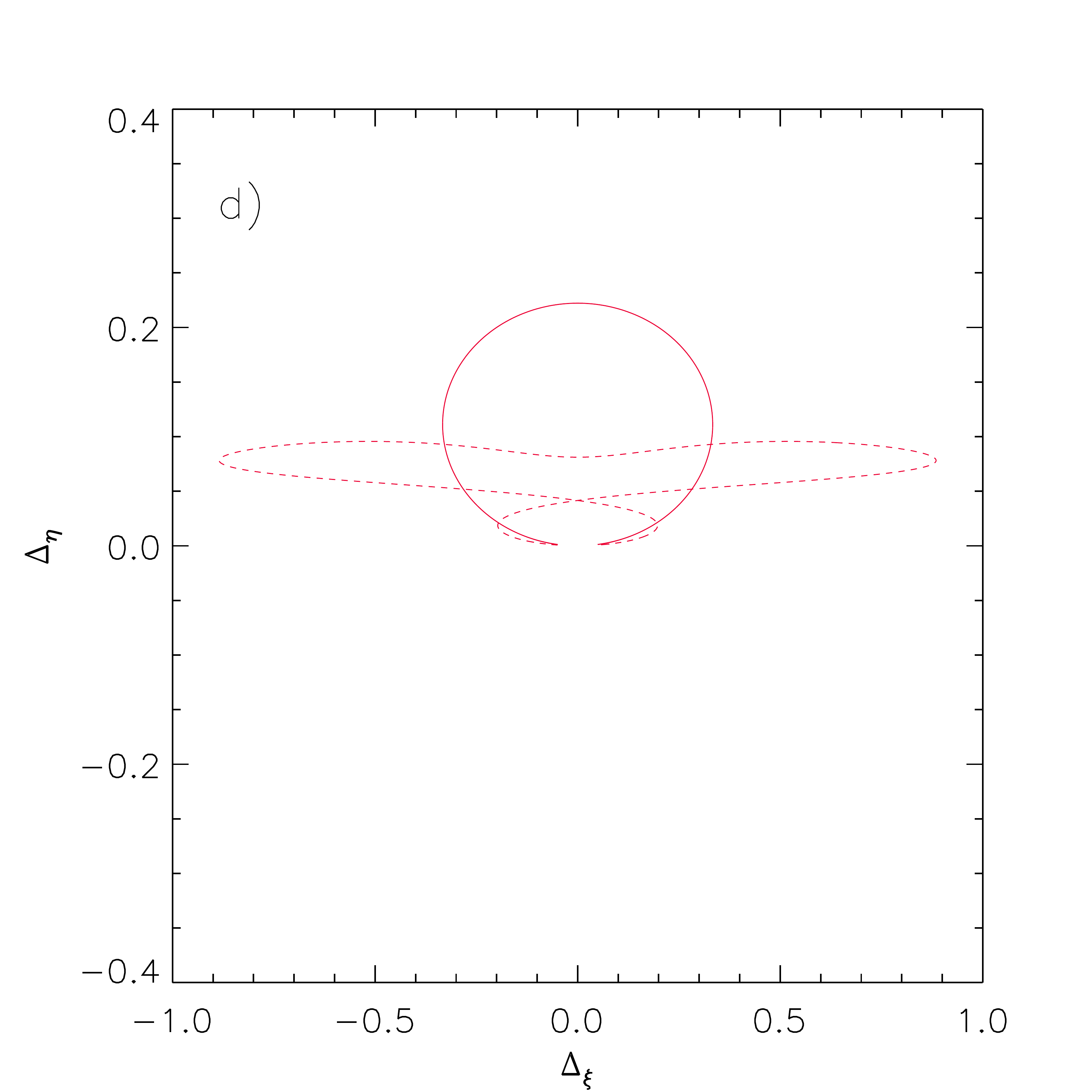}
        }
 \caption{We give the source path (left panels) and the corresponding astrometric curves (right panels) 
for a binary source with components of mass $m_1=m_2=1$ M$_{\odot}$, with the same luminosity, and separated by $10$ AU. 
The lens has a mass of $m_l=1$ M$_{\odot}$, is located at $D_L=1$ kpc and moves with a projected velocity $v_{\perp}=100$ km s$^{-1}$, thus implying an Einstein angular radius of $2.7$ mas. 
The event impact parameter is $u_0=0.5$. The upper panels show the expected signal for a static binary source, 
while in the bottom ones the orbital motion is taken into account. For the simulated event, the Einstein time and the 
binary source orbital period are $t_E\simeq 46$ days and $P\simeq 8674$ days, respectively. In the right panels the continuous ellipses represent the astrometric trajectories 
for a single source located in the center of mass, while dashed lines stand for the binary source.}
   \label{fig1paper2}
\end{figure*}

\begin{figure*}
 \centering
        \subfigure{%
            \includegraphics[width=0.9\columnwidth]{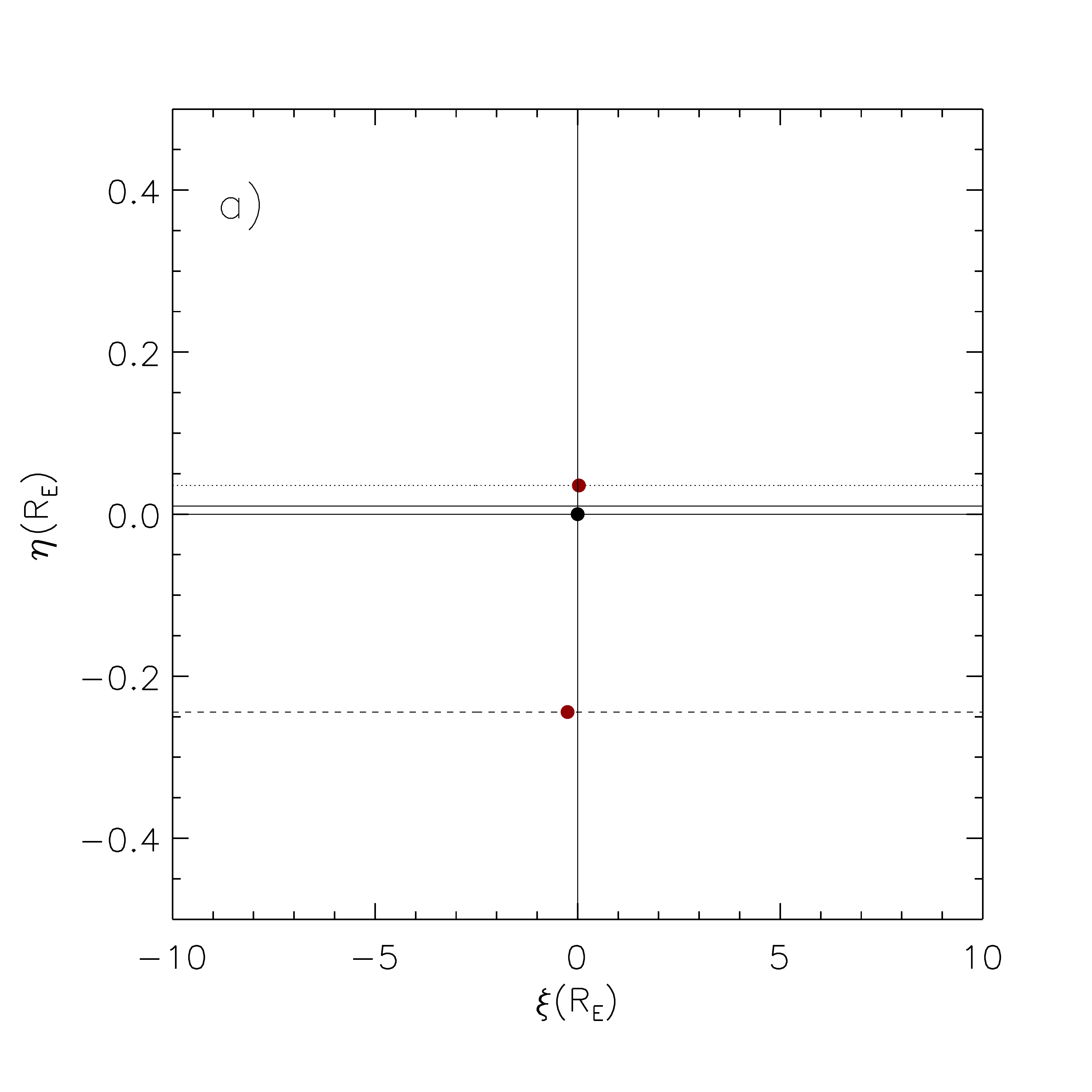}
        }
        \subfigure{%
           \includegraphics[width=0.9\columnwidth]{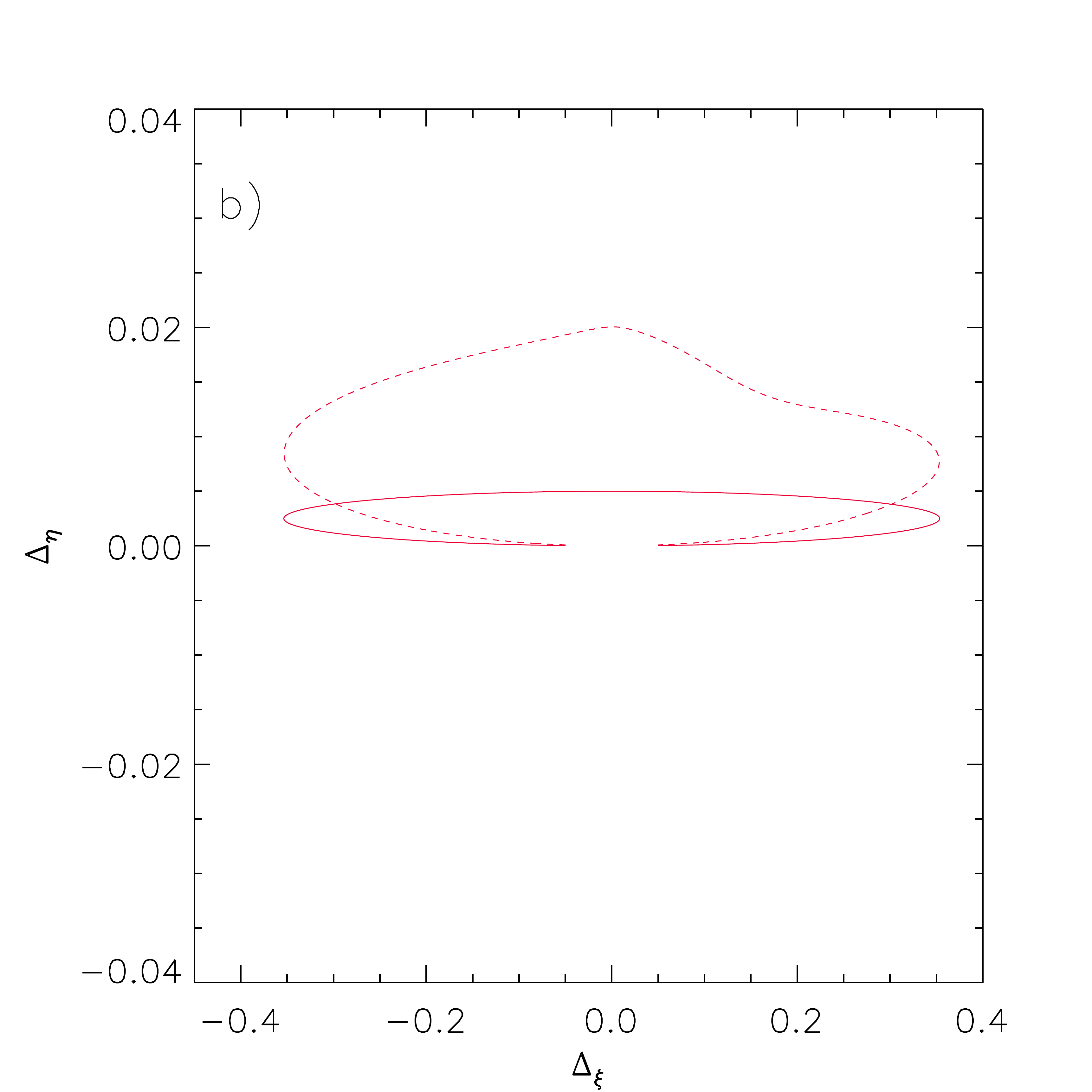}
        }\\
        \subfigure{%
            \includegraphics[width=0.9\columnwidth]{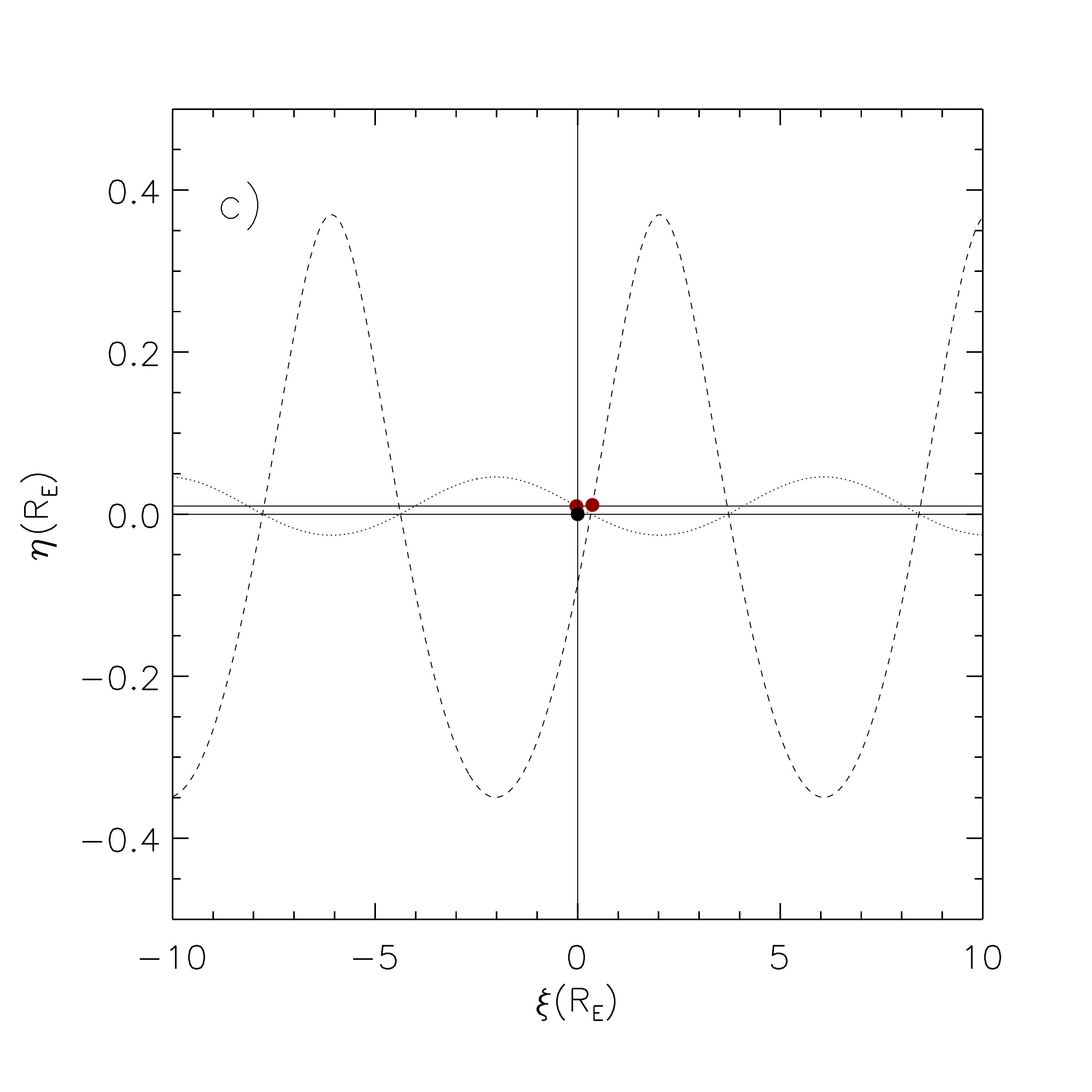}
        }%
        \subfigure{%
            \includegraphics[width=0.9\columnwidth]{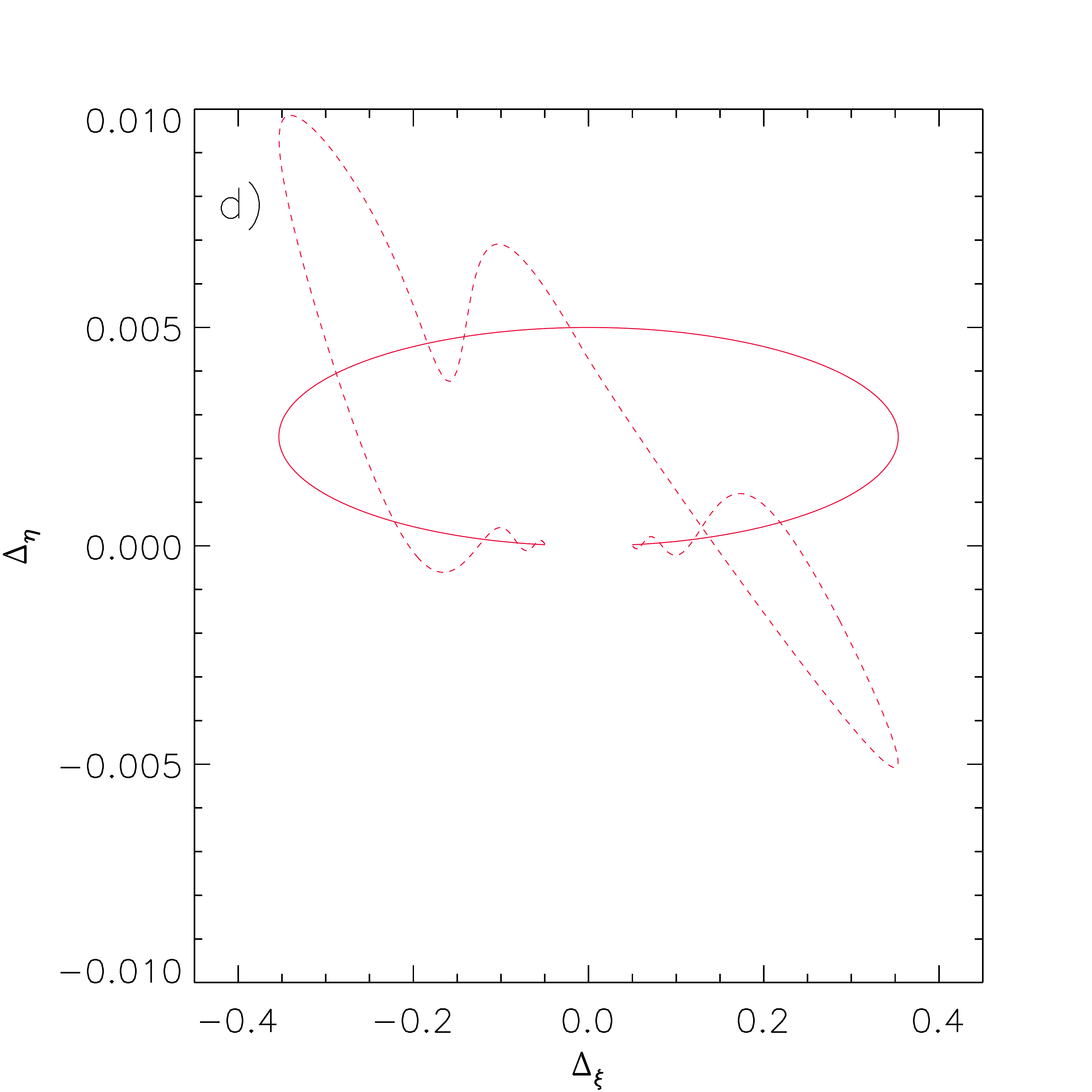}
        }
 \caption{The same as in Figure 1, for a binary source with components of mass $m_1=1$ M$_{\odot}$, $m_2=0.1$ M$_{\odot}$, $F_1=1$, $F_2=10^{-2}$, 
and separated by $1$ AU. The lens has a mass of $m_l=1$ M$_{\odot}$, is located at $D_L=1$ kpc 
and moves with a projected velocity $v_{\perp}=100$ km s$^{-1}$, thus implying an Einstein angular radius of $2.7$ mas. 
The event impact parameter is $u_0=0.01$. For the simulated event, the Einstein time and the 
binary source orbital period are $\simeq 46$ days and $\simeq 369$ days, respectively.}
   \label{fig2paper2}
\end{figure*}

In Figure \ref{fig1paper2}, we {  present} the source path (left panels) and the astrometric shift (right panels) for a simulated microlensing event involving a binary source. The binary source system is 
constituted by two objects with equal mass ($m_1=m_2=1$ M$_{\odot}$) and luminosity ($F_1=F_2=1$ L$_{\odot}$), separated by a distance of $10$ AU. The binary source is assumed to reside in the galactic 
bulge, i.e. $D_S\simeq 8$ kpc. The lens (located at $D_L=1$ kpc from the observer) has mass $m_l=1$ M$_{\odot}$, impact parameter $u_0=0.5$, and moves with a projected velocity $v_{\perp}=100$ km s$^{-1}$, 
thus implying an Einstein angular radius $\theta_E=2.7$ mas. For the simulated event, $t_E \simeq 46$ days and $P\simeq 8674$ days.
In panels (a) and (b), we consider {  a static} binary 
source, while in panels (c) and (d) the source system orbital motion is taken into account. In Figure \ref{fig2paper2}, we consider the expected astrometric microlensing signal for a static -- panels (a) and (b) -- and rotating -- panels (c) and (d) -- binary source, respectively. Here, we assumed two objects 
with masses $m_1=1$ M$_{\odot}$, and $m_2=0.1$ M$_{\odot}$, separated by $1$ AU and fixed the intrinsic luminosities to $F_1=1$ L$_{\odot}$, and $F_2=0.01$ L$_{\odot}$. We furthermore set $u_0=0.01$. 
For such case, the binary source orbital period turns out to be $P\simeq 370$ days.
{  In both Figures}, the solid curves represent the centroid shift ellipse expected for a single source located {  at} the center 
of mass of the binary source system. It is evident that the presence of a binary source system
introduces deformations of the astrometric signal with respect to the pure ellipse case. 
{  This is also true when the orbital motion of the binary source system is taken into account as illustrated in the lower panels of 
Figures \ref{fig1paper2} and \ref{fig2paper2}, where modulations 
with a time scale corresponding to the source system orbital period do appear.}
  
Note that, for the considered cases, being $\theta_E\simeq 2.7$ mas, the astrometric signal results well within the astrometric 
precision of the Gaia satellite in five years of integration. This opens the possibility to detect binary systems as sources 
of astrometric microlensing events and characterize their physical parameters (mass ratio, projected separation and orbital period).

We would like to mention the challenging possibility for Gaia-like observatories to detect also astrometric microlensing 
events involving both binary sources and binary lenses. For the sake of simplicity, we do not consider here the orbital motion of the systems. 
For such cases, eq. (\ref{deltatotalbs}) remains valid provided that the centroid shifts ${\rm \Delta_i}$ of each components of the binary source system are obtained 
solving numerically the two body lens equation. In this case the lens equation {  is expressed as} (\citealt{witt1990,witt1995,sgnumerical}),
\begin{equation}
\zeta_i=z_i+\frac{m_{L,1}}{z_{L,1}+\bar{z}_i}+\frac{m_{L,2}}{z_{L,2}+\bar{z}_i},  
\end{equation}
where $m_{L,1}$ and $m_{L,2}$ are the masses of the binary lens components (with  $m_{L,2}< m_{L,1}$ so that $q_L=m_{L,2}/m_{L,1}<1$ ), $z_{L,1}$ and 
$z_{L,2}$ the positions of the lenses (separated by $b_L$), and $\zeta_i=\xi_i+i\eta_i$ and $z_i=x_i+iy_i$ the positions of the binary source components and 
associated images, respectively. The lens components are located on the $\xi$ axis with the primary at $(-b_L/2,0)$ and secondary at $(b_L/2,0)$. 
\begin{figure*}
 \centering
        \subfigure{%
            \includegraphics[width=0.9\columnwidth]{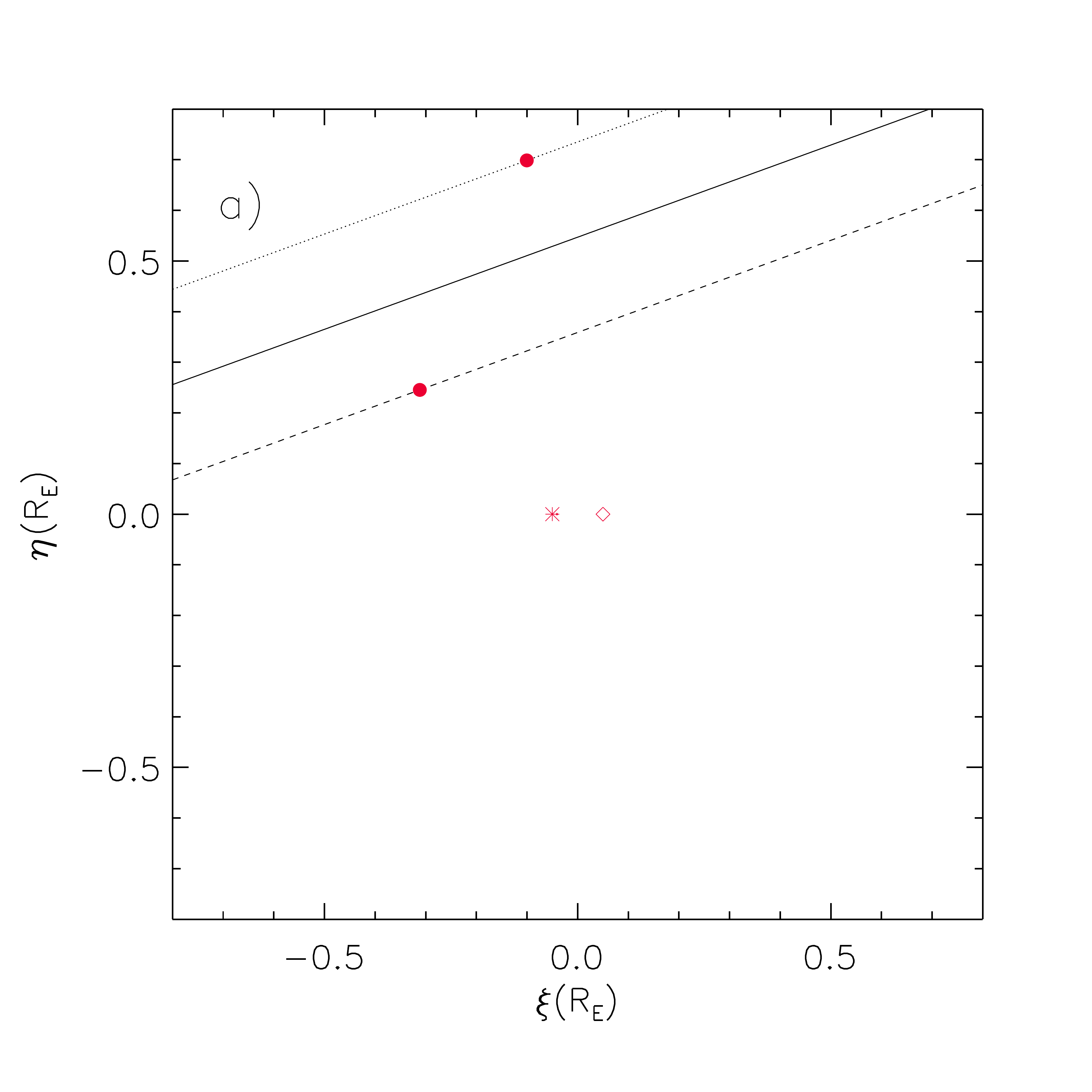}
        }
        \subfigure{%
           \includegraphics[width=0.9\columnwidth]{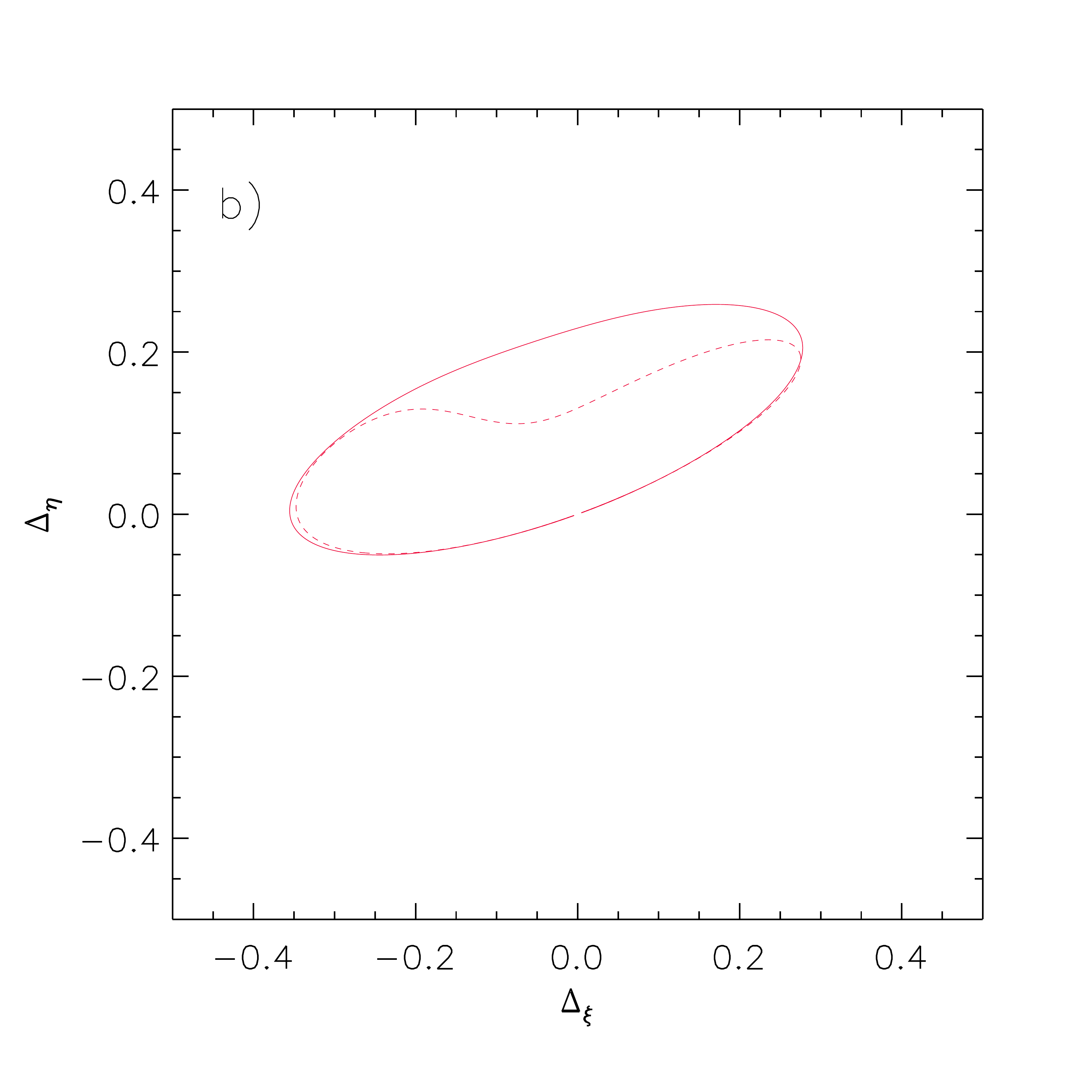}
        }\\
        \subfigure{%
            \includegraphics[width=0.9\columnwidth]{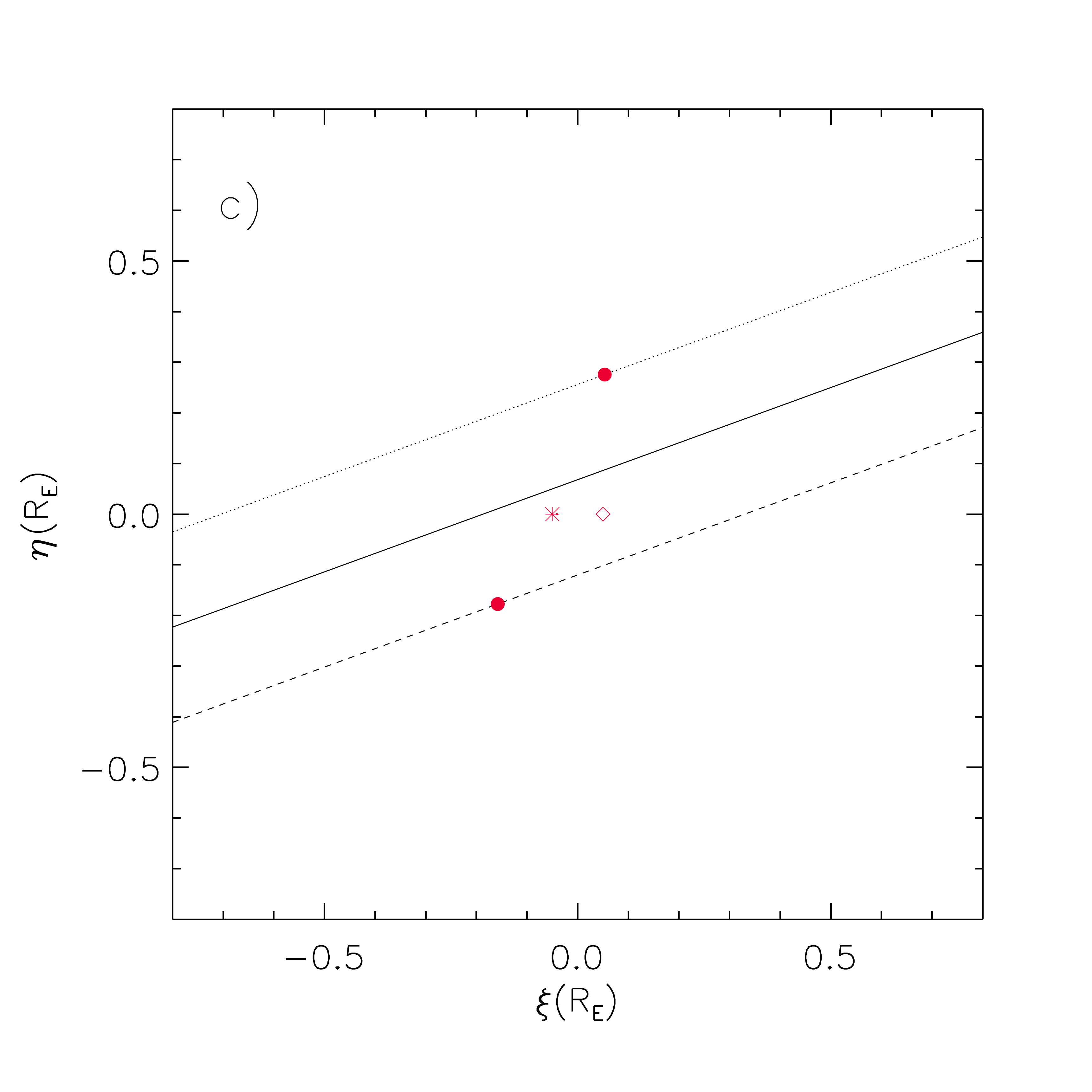}
        }%
        \subfigure{%
            \includegraphics[width=0.9\columnwidth]{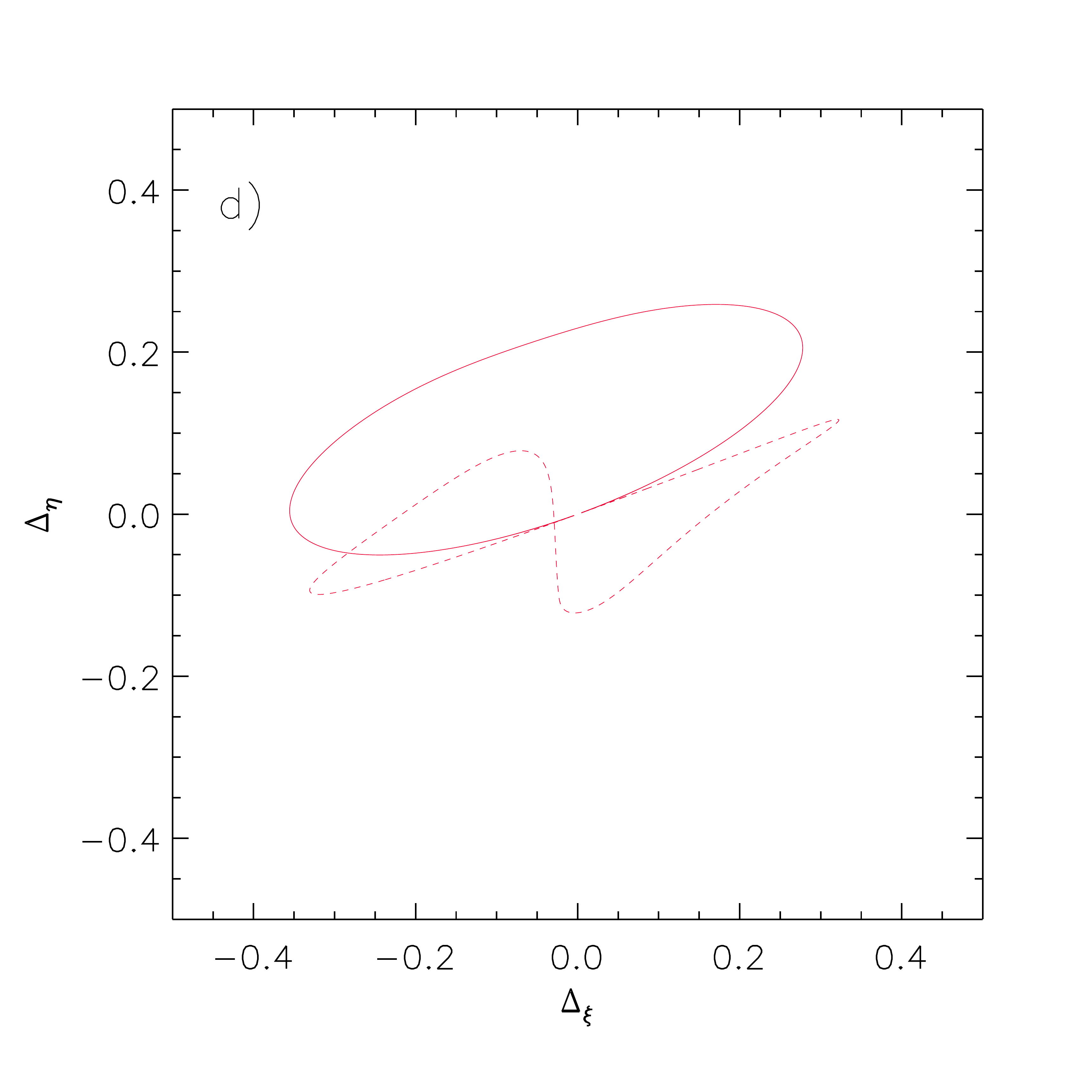}
        }
 \caption{We give the source path (left panels) for the primary source (dotted line), its companion (dashed line), and the associated center of mass (solid line). The asterisk and diamond represent 
the location of the primary and secondary lens, resepctively.
The expected astrometric signal is given in the right panels 
for a binary source ($b=0.5$, $q=1$) lensed by a binary lens ($b=q=0.1$) with impact parameter $u_0=0.5$ (panel b) and impact paremeter $u_0=0.05$ (panel d).}
   \label{fig3paper2}
\end{figure*}
\begin{figure*}
 \centering
        \subfigure{%
            \includegraphics[width=0.9\columnwidth]{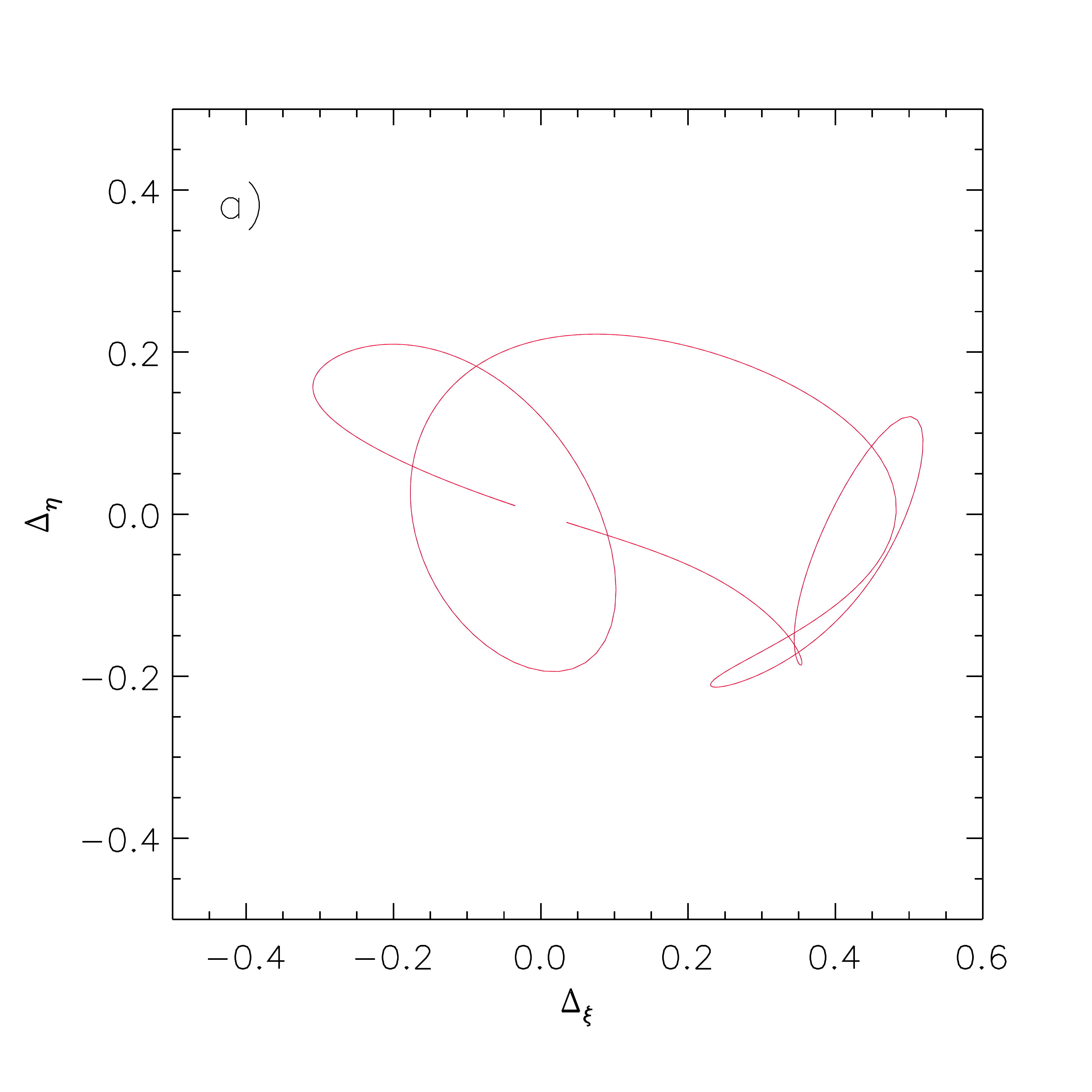}
        }
        \subfigure{%
           \includegraphics[width=0.9\columnwidth]{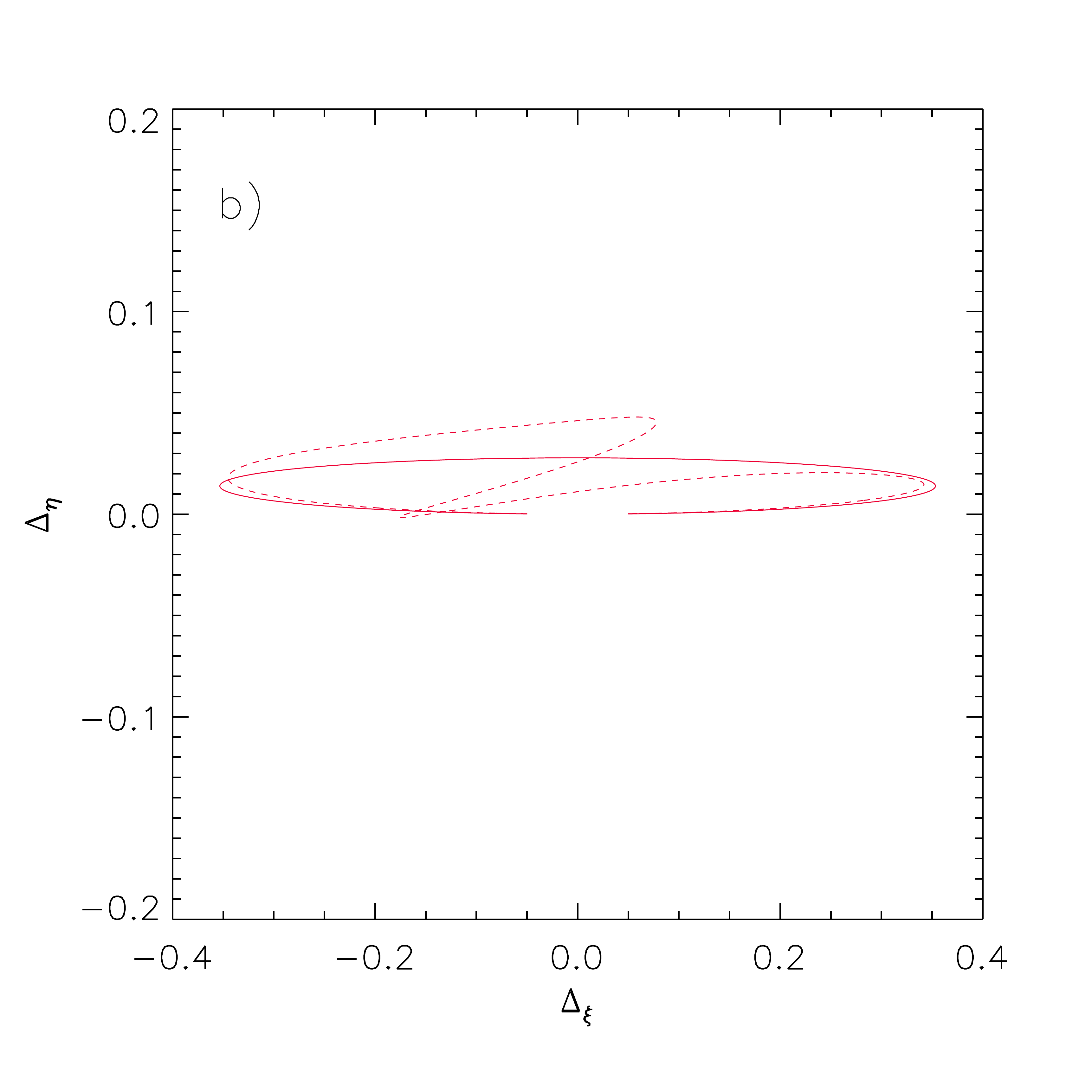}
        }
 \caption{For the event OGLE 2002-BLG-099, we give the simulated astrometric path for the binary lens and single source (left panel for $q=0.248$, $b=1.963$, $u_0=0.09$, and $t_E=24.4$ days) and for the single lens and binary source (right panel 
for $u_{0,1}=0.0821$, $u_{0,2}=0.0294$, corresponding to peak 
times of  $t_{0,1}=2402.93$ days and $t_{0,2}=2425.23$, $t_E=47.1$ days, and blending parameters $f_1=0.147$ and $f_2=0.051$) cases, respectively.}
   \label{fig4paper2}
\end{figure*}

In this case, the centroid shifts with respect to the position of the unlensed star (one per each of the intervening source, see also  
\citealt{han2001astrometry}) are  
\begin{equation}
(\Delta_{\xi,i}, \Delta_{\eta,i})=(x_{c,i}-\xi_i, y_{c,i}-\eta_i),
\end{equation}
where the positions of the source star centroid are simply the average of the locations of the individual images wighted by each corresponding amplification $\mu_{j,i}$, i.e.
\begin{equation}
(x_{c,i}, y_{c,i})=\left(\sum_j\mu_{j,i}x_i/\mu_i, \sum_j\mu_{j,i}y_i/\mu_i\right).
\end{equation}
Here, $\mu_i$ is the total amplification (i.e. $\mu_i=\sum_j \mu_{j,i}$, with $j$ running over the image number) and, as above, $i=1,2$ indicates the primary and secondary component of the binary source system. As an example, 
in panel (a) of Figure \ref{fig3paper2}, we {  present the paths of the primary} (dotted line), and secondary (dashed line) components of the binary source system characterized by $b=0.5$, and $q=1$. 
The solid line indicates the path followed by the center of mass. The binary source system (assumed here for simplicity to be static) is lensed  by a binary lens with $b_L=q_L=0.1$ and  the event impact parameter is
$u_0=0.5$. The asterisk and diamond represent the position of the primary and secondary lenses, respectively. In panel (b) we give the astrometric signal 
(dashed line) expected for the simulated microlensing event. For comparison, the solid line represents the astrometric signal associated to 
the same binary lens acting on a single point-like source. Note that the presence of a binary source gives a substantial difference with respect to the single source case that, for the assumed simulated event parameters, 
amounts to $\simeq 0.1\theta_E\simeq 270$ $\mu$as, well within the Gaia capabilities.
In panels (c) and (d), we set the event impact parameter to $u_0=0.05$, leaving the other parameters unchanged. In this case the astrometric signal is completely 
different with respect to the previous case. This is a general behaviour of the astrometric shift curves which strongly depend even to small changes of the system physical parameters.
It goes without saying that, conversely to what happens with the standard photometric microlensing, a {  fitting} procedure on the observed astrometric 
data may provide a robust estimate of the microlensing event parameters. 
This is clear when considering {  events not well sampled as  
OGLE 2002-BLG-099 (see \citealt{jaroszynski2004} for details) 
where different interpretations of the photometric data are statistically acceptable}. In particular, the considered event can be described as due to a single source 
lensed by a binary system or by a double source lensed by a single object. While the light curve analysis does not allow one to distinguish between these models, it is clear from Fig. \ref{fig4paper2}, 
that astrometric observations would have resolved the degeneracy since the astrometric signals associated to the two cases are completely different. Indeed, considering the most likely values for 
the total lens mass and distance of $\simeq 0.47$ M$_{\odot}$ and $D_L\simeq 5.7$ kpc (\citealt{dominik2006}) one gets $\theta_E\simeq 470$ $\mu$as. Hence, from Fig. \ref{fig4paper2}, astrometric 
observations with precision of at least $40-50$ $\mu$as (i.e. within the capabilities of the Gaia satellite) would make possible the distinction between the two different configurations.

\section{Earth parallax effects on astrometric microlensing}

In photometric observation of microlensing events the parallax effect, due to the Earth motion, 
generally induces minor anomalies unless the event Einstein time is comparable with (or longer than) the Earth orbital period (see, e.g., \citealt{Wyrzykowski2016}). 
On the contrary, in astrometric microlensing the Earth orbital motion is not negligible even for short duration events. {  Here, based on the seminal idea by \citet{paczinsky1998} 
we account for the parallax effect following}
\begin{figure}
      \centering
        \subfigure{%
            \includegraphics[width=0.85\columnwidth]{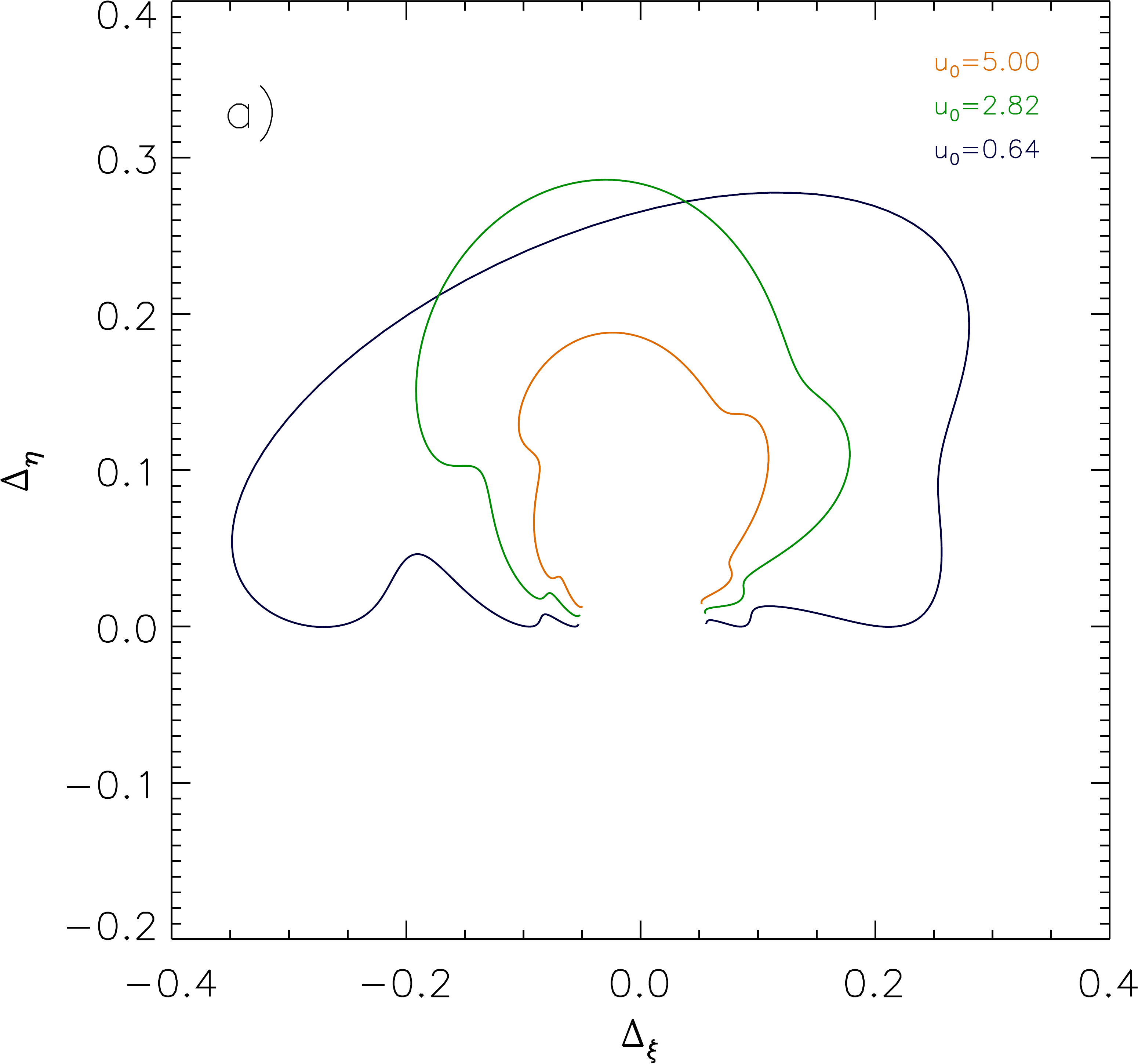}
        }\\
        \subfigure{%
           \includegraphics[width=0.85\columnwidth]{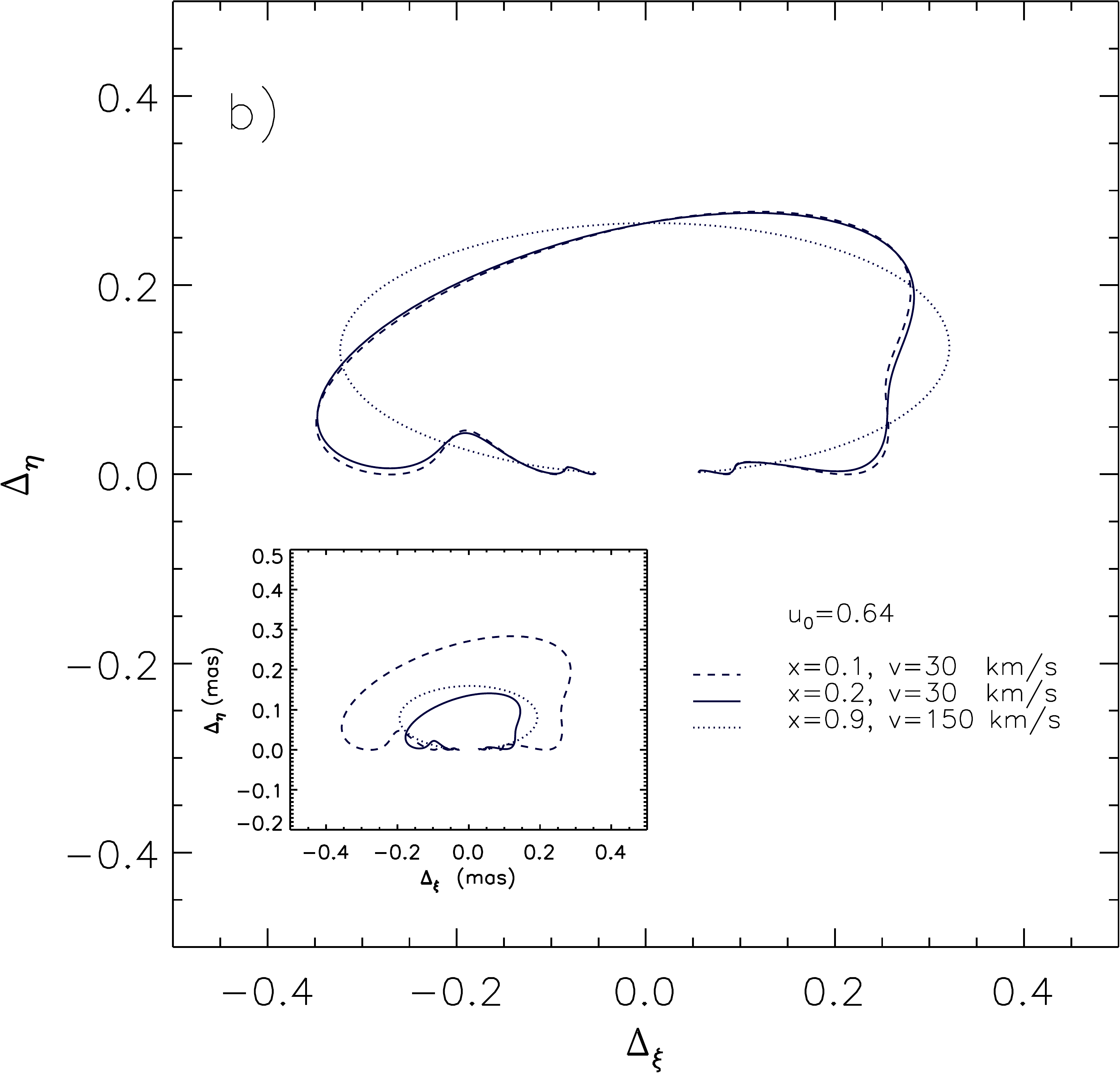}
        }\\
        \subfigure{%
            \includegraphics[width=0.85\columnwidth]{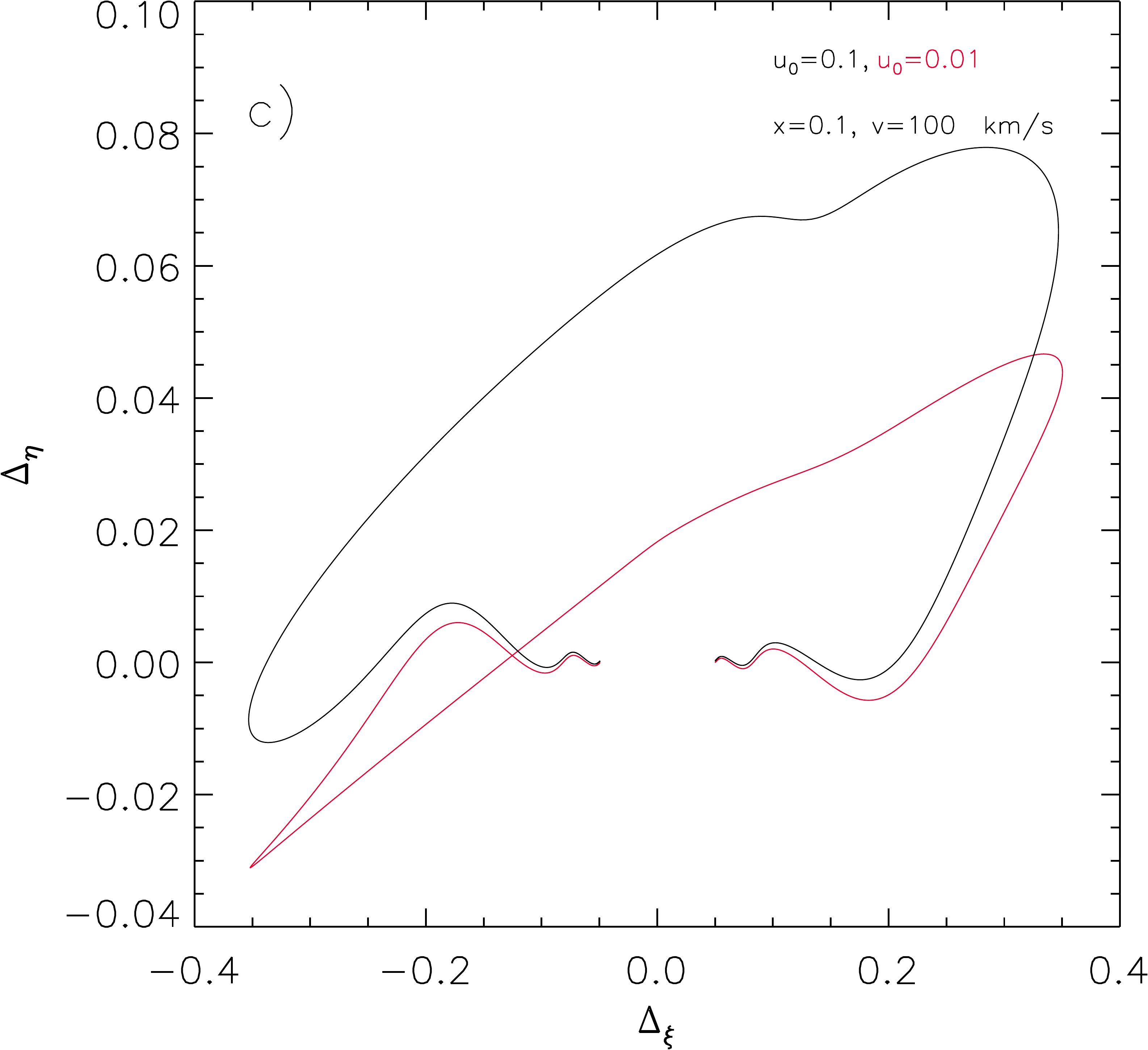}
        }
   \caption{Earth parallax effect on astrometric curves for three simulated microlensing events (see text for details).}
   \label{parallasse}
\end{figure}
the formalism provided by \citet{dominik1998a} in the approximation of small orbital eccentricity.  Let \(\xi_{0}(t)\) and \(\eta_{0}(t)\) be the coordinates
of the source (not corrected for parallax effect) in the lens plane at time \(t\). The new coordinates are
\begin{subequations}
  \begin{align}
    \begin{split}
      \xi(t) &= \xi_{0}(t) + (\tilde{x}_{1}(t) - \tilde{x}_{1}(t_{0}))\cos\psi\\
           &\quad{}+ (\tilde{x}_{2}(t) - \tilde{x}_{2}(t_{0}))\sin\psi,
   \end{split} \\
    \begin{split}
      \eta(t) &= \eta_{0}(t) - (\tilde{x}_{1}(t) - \tilde{x}_{1}(t_{0}))\sin\psi\\
           &\quad{}+ (\tilde{x}_{2}(t) - \tilde{x}_{2}(t_{0}))\sin\psi,
    \end{split}
  \end{align}
\end{subequations}
where
{\allowdisplaybreaks[4]
\begin{subequations}
  \begin{align}
    \tilde{x}_{1}(t) &= -A'(t)\sin\chi\cos\nu(t), \\
    \tilde{x}_{2}(t) &= A'(t)\sin\nu(t), \\
    A'(t) &= a_{\oplus}\bigl[1 - \varepsilon_{\oplus}\cos M(t)\bigr]
            \frac{1 - x}{R_{\textup{E}}}, \\
    \nu(t) &= M(t) + 2\varepsilon_{\oplus}\sin M(t) - \varphi, \\
    M(t) &= 2\pi\frac{t - t_{\textup{p}}}{P_{\oplus}}, \\
    x &=  D_{\textup{L}}/D_{\textup{S}}.
  \end{align}
\end{subequations}}
In the previous equations, \(M(t)\) is the mean anomaly of the Earth, \(t_{\textup{p}}\) is the last time
of perihelion passage, so that \(M\) lies in the interval \([0, 2\pi[\)\,,
\(\nu(t)\) its true anomaly shifted by \(\varphi\).  In addition, \(\varphi\)
and \(\chi\) are, respectively, the longitude and the latitude of the source
measured in the ecliptic plane as prescribed by \citet{dominik1998a}, while \(\psi\) is the
relative orientation of \(\boldsymbol{v}_{\perp}\) to the Sun-Earth system.
Here, \(a_{\oplus} \simeq 1.49\times 10^{13}~\textup{cm}\) is the Earth orbit semi-major
axis\footnote{Note that the Gaia satellite is placed at the Lagrangian Point L2 at about $1.5\times 10^{6}$ km from Earth, a distance much smaller 
than the Sun-Earth semi-major axis. It is therefore reasonable to apply the Earth parallax correction 
described in this Section also to Gaia observations.}, \(\varepsilon_{\oplus} = 0.0167\) is its eccentricity, and
\(P_{\oplus} = 365.26~\textup{d}\) the orbital period.  In the definition of
\(A'(t)\), note that
\(\rho' = a_{\oplus}(1 - x)/R_{\textup{E}}\) is the
Earth semi-major axis projected onto the lens plane, in units of Einstein
radii, and is a measure of the importance of parallax effect. 

In Figure \ref{parallasse}, we show the astrometric curves (obtained with an integration time of $\simeq 5$ years) for three simulated events taking into 
account the Earth parallax and assuming $t_0=t_p=0$. In all cases, we fixed the source coordinates to be $\varphi = 2.93$ rad, and $\chi=-0.08$ rad as in 
\citet{dominik1998a} cooresponding to ecliptic coordinates $\lambda=271\degr$, and $\beta=-5\degr$.
 
In the upper panel, a single source is microlensed by a single lens ($x=0.1$, $v=30$ km s$^{-1}$ and $t_E=50$ days corresponding to $\theta_E\simeq 1$ mas) 
for three different impact parameters. In the middle panel, we fixed the impact parameter to $u_0=0.64$ leaving the other parameters unchanged. 
Dashed and continuous curves are for two disk events at different distances from the observer, 
$x=0.1$ and $x=0.2$ (corresponding to $\theta_E\simeq 1$ mas and  $\theta_E\simeq 0.5$ mas), respectively. The dotted line has been obtained for a bulge lens ($x=0.9$) with 
$\theta_E\simeq 0.6$ mas. The inset shows the scaled astrometric signal in physical units.      
Finally, in the bottom panel, we give the expected astrometric signal (red curve) for a (static) binary source microlensed by a single object, assuming the   
same parameters as in Figure \ref{fig2paper2} and $u_0=0.01$. The black line corresponds to an event with impact parameter $u_0=0.1$. In both cases, $x=0.1$, 
$\theta_E\simeq 2.5$ mas and $t_E=45.6$ days.

It is worth mentioning that while in standard photometric microlensing the parallax effect becomes more important close to the event peak and (especially) for long events, 
in astrometric observations the deviations with respect to the pure ellipse path show up even in the case of events characterized by short $t_E$. Moreover, modulations with the Earth orbital period 
appear, also at very large impact parameter values where the photometric signal is useless. As a final note, we remark that taking into account the source orbital motion produces 
modulations with a peculiar frequency characteristic of the system.

\section{Conclusions}

In this paper we considered the anomalies induced in simulated astrometric events by the orbital motion of the lens and/or source binary systems taking into account the Earth parallax effect. 
Considering and implementing these effects in astrometric microlensing is essential in order to correctly estimate the system parameters thus alleviating the parameter degeneracy problem that 
afflicts photometric microlensing. This issue is particularly important in the era of Gaia satellite that is performing a survey of the whole sky
allowing one to get the astrometric path of microlensed sources with unprecedent precision. Indeed, it has been
estimated that the Gaia mission will discover, in five years of operations, $\simeq 3500$ photometric and $25000$ astrometric microlensing events (\citealt{Belokurov2002}) 
which will be characterized by a astrometric precision down to 30 $\mu$as. An even better precision could be possibly obtained by following up the events discovered by Gaia 
with ground-based observations (as Gravity at VLT, see \citealt{eisenhauer2009} but also \citealt{zurlo2014}) for a longer observation time. 
This is important since the astrometric path of microlensing event changes substantially during a much longer time interval than in the usual  photometric observations.

\section*{Acknowledgments}
{We acknowledge the support by the INFN project TAsP (Theoretical Astroparticle Physics Project). 
MG would like to thank Max Planck Institute for Astronomy (Heidelberg), where part of this work has been done, and Luigi Mancini for hospitality. 
We also thank the anonymous referee for the constructive comments.}

\label{lastpage}

\end{document}